\documentclass[12pt]{article}
\usepackage{epsfig}
\usepackage{amssymb}

\newcommand{\be}{\begin{equation}}
\newcommand{\ee}{\end{equation}}
\newcommand{\bea}{\begin{eqnarray}}
\newcommand{\eea}{\end{eqnarray}}
\newcommand{\ov}{\overline}

\newcommand{\eps}{\epsilon}
\newcommand{\ba}{\begin{array}}
\newcommand{\ea}{\end{array}}

\begin{document}

\vspace{10mm}
\centerline{\LARGE\bf Gluon condensate and {\boldmath $c$}-quark mass in}
\vspace{3mm}
\centerline{\LARGE\bf pseudoscalar sum rules at 3-loop order}
\vspace{10mm}
\centerline{\large K.N. Zyablyuk}
\vspace{3mm}
\centerline{\tt zyablyuk@heron.itep.ru}
\vspace{5mm}
\centerline{\it Institute of Theoretical and Experimental Physics,}
\centerline{\it B.Cheremushkinskaya 25, Moscow 117218, Russia}

\begin{abstract}
Charmonium sum rules for pseudoscalar $0^{-+}$ state $\eta_c(1S)$ are analyzed
within perturbative QCD and Operator Product Expansion. The perturbative
part of the pseudoscalar correlator is considered at $\alpha_s^2$ order and the
contribution of the gluon condensate $\left<G^2\right>$ is taken into account with
$\alpha_s$ correction. The OPE series includes the operators of dimension $D=6,8$ 
computed both in the instanton and factorization model.
The method of moments in $\ov{\rm MS}$ scheme allows
to establish acceptable values of the charm quark mass and gluon condensate,
using the experimental mass of $\eta_c$. In result of 
the analisys the charm quark mass is found to be ${\bar m}_c=1.26\pm 0.02\,{\rm GeV}$ 
independently of the condensate value.
The sensitivity of the results to various approximations for the massive 3-loop
pseudoscalar correlator is discussed.
\end{abstract}

\centerline{PACS: 13.35.D, 11.55.H, 12.38}


\section{Introduction}

The concept of Operator Product Expansion (OPE) was applied to QCD sum rules in \cite{SVZ} to
parametrize the nonperturbative effects. The operators of increasing dimension,
constructed from quark and gluon fields, or condensates, constitute the OPE series, which
is added to the perturbative ones. In case of heavy quark correlators the quark condensates
are not essential and OPE series start from the dimension $D=4$ gluon condensate
$$
\left< {\alpha_s\over \pi} G^a_{\mu\nu} G^a_{\mu\nu} \right>
$$
for which the authors of \cite{SVZ} have obtained the estimation $0.012\,{\rm GeV}^4$ from
vector charmonium sum rules. In \cite{SVVZ} the charmonium sum rules were
studied in pseudoscalar channel and it was predicted
the mass of the lowest $\eta_c$ state $3.00\pm 0.03 \,{\rm GeV}$. This result
was in contradiction with available to that time experimental information. Later measurements
found the mass of $\eta_c$ close to $3.0\,{\rm GeV}$, which was considered as a triumph
of QCD.

Since then various sum rules were analyzed in many publications\footnote{See \cite{IZ}
for the list of publications} in order to obtain or specify the value of the gluon condensate.
In the recent paper \cite{IZ} the vector charmonium sum rules were reconsidered with
$\alpha_s^2$-corrections of the perturbative series
and $\alpha_s$-corrections to the condensate contribution,
 and up to date experimental data. The analysis
of \cite{IZ} resulted to the gluon condensate $0.009\pm 0.007 \, {\rm GeV}^4$
and $c$-quark mass $m_c(m_c)=1.275\pm 0.015 \,{\rm GeV}$.
Despite high accuracy of experimental data and $c$-quark mass determination,
 the accuracy of the gluon condensate remains $\sim \pm 100\%$, and zero value is not excluded.

It also seems reasonable to reanalyze the pseudoscalar sum rule, taking into account the
information obtained in \cite{IZ}. Now the mass of $\eta_c$ is known experimentally with high accuracy
$2979.7\pm 1.5 \,{\rm MeV}$ \cite{PDG}, so we invert the problem and find a restriction on
the charm quark mass and gluon condensate, imposed by this sum rule. A special attention
should be paid to the correlation between these two values, since a variation of one
parameter leads to the change of another.

The sum rule technique goes as follows.
The correlator of the pseudoscalar charm currents is defined as:
\be
\Pi^p(q^2)\,=\,i\int dx\, e^{iqx} \left< \,TJ^p (x) J^p (0) \, \right> \; , \qquad J^p = 2 i m\, \bar{c} \gamma_5 c
\label{popdef}
\ee
We define the pseudoscalar current as $J^p=\partial_\mu J^a_\mu$,
 $J^a_\mu={\bar c}\gamma_5\gamma_\mu c$
is axial vector current. Within the narrow width approximation the imaginary part is:
\be
 {\rm Im} \, \Pi^p(q^2+i0)\,=\,\pi\sum_\eta \, \delta(q^2-m_\eta^2) \, \left|\left<0|J^p(0)|\eta\right>\right|^2
\label{impop}
\ee
The sum goes over the pseudoscalar states\footnote{The next to lightest pseudoscalar
state $\eta_c(2S)$ with mass $3654 \pm 6(stat)\pm 8(syst)\,{\rm MeV}$ was discovered recently \cite{Belle}}
 with $J^{PC}=0^{-+}$. The correlator (\ref{popdef}) is quadratically divergent, so the
dispersion relation requires double subtraction:
\be
\label{drpsc}
\Pi^p(q^2)\,=\,c_0\,+\,q^2 c_1 \,+\,
{q^4\over \pi} \int_0^\infty {{\rm Im}\,\Pi^p(s+i0)\over s^2( s-q^2)} \,ds
\ee
provided the integral in the rhs is convergent, $c_0$, $c_1$ are unknown constants.
In order to suppress the contribution of the higher states in (\ref{impop}) as well as continuum contribution,
one considers the derivatives of the polarization operator
in euclidean region $Q^2\equiv -q^2>0$, the so-called moments:
\bea
M_n^p(Q^2) &  \equiv & {8\pi^2 \over 3 n!} \left( - {d\over dQ^2} \right)^n \Pi^p(Q^2) =
 {8\pi\over 3} \int_0^\infty {{\rm Im}\,\Pi^p(s+i0)\over (s+Q^2)^{n+1}}\,ds\nonumber \\
 & & \,=\,
{8\pi^2\over 3}\sum_\eta\,{\left|\left<0\left|J^p(0)\right|\eta\right>\right|^2\over (m_\eta^2+Q^2)^{n+1}}
\label{mompdef}
\eea
where $n\ge 2$. The matrix elements $\left<0\left|J^p(0)\right|\eta\right>$ are not known experimentally.
But if one considers the ratio of some two moments at sufficiently high $n$, the contribution
of the lightest state $\eta_c(1S)$ becomes dominant and
\be
\label{raps0}
{M_n^p(Q^2)\over M_{n+1}^p (Q^2)}\,=\,m_{\eta_c}^2+Q^2 \; , \qquad n\to \infty
\ee
This property was exploited to predict the mass of $\eta_c$ in \cite{SVVZ, RRY}.
An essential point was noticed in \cite{RRY}: the QCD corrections to the moments are large at
$Q^2=0$, so the sum rules should be considered at $Q^2>0$. Moreover, huge contribution of
the dimension 8 operators $\left<G^4\right>$ to the moments at $Q^2=0$ \cite{NR2} becomes
tolerable at $Q^2\sim 4m_c^2$.

The subject of this paper is a detailed analysis of the sum rule (\ref{raps0}). In the next
section the perturbative and OPE corrections to the correlator (\ref{popdef}) are described.
Section 2 is devoted to the moments both in the pole and $\ov{\rm MS}$ scheme for the charm quark
mass. In the Section 3 various contribution to the pseudoscalar sum rule (\ref{raps0}) 
are studied in details for typical values of the charm mass and gluon condensate. 
The higher dimension $D=6,8$ gluon operators are calculated both in the instanton and 
factorization model. In the final Section the restriction on the $c$-quark mass 
and the gluon condensate $\left< aG^2 \right>$ are obtained.


 \section{Pseudoscalar correlator in  QCD}

In QCD the polarization function (\ref{popdef}) consists of perturbative part and
operator product expansion:
\be
\label{ppptope}
\Pi^p\,=\,\Pi^p_{\rm PT} \,+\,\Pi^p_{\rm OPE}
\ee
The perturbative part is determined by its imaginary part via dispersion relation (\ref{drpsc}).
The imaginary part is parametrized by the coefficient functions $R^{(k),p}$ in the
expansion by the running QCD coupling $a(\mu^2)\equiv \alpha_s(\mu^2)/\pi$:
\be
\label{rpcf}
{\rm Im}\,\Pi^p_{\rm PT}(s+i0)\,=\,{3\,sm^2\over 2\pi}\, \sum_{k\ge 0} R^{(k),p}(s,\mu^2)\, a^k(\mu^2)
\ee
It is simpler to parametrize the functions $R^{(k),p}$ in terms of
the pole masse $m$ of $c$-quark.
The first two terms do not depend on the scale $\mu^2$. They are known analytically \cite{HT}:
\bea
R^{(0),p} & = &  v  \nonumber \\
R^{(1),p} & = & {v\over 2}(7-v^2)+4v\left( \ln{1-v^2\over 4} -{4\over 3} \ln{v} \right) + {19+2v^2+3v^4\over 12}
\ln{1+v\over 1-v} +{8\over 3}(1+v^2) \nonumber \\
 & &
 \times \left[ \,2\,{\rm Li}_2\!\left( {1-v\over 1+v} \right) +\,{\rm Li}_2\!\left( -{1-v\over 1+v} \right)
+ \left( {3\over 2}\ln{1+v\over 2}- \ln{v} \right) \ln{1+v\over 1-v}\,\right]
\label{rpcf12}
\eea
Here and below $v=\sqrt{1-1/z}$, $z=s/(4m^2)$.
 The function $R^{(2),p}$ is usually decomposed into the following gauge invariant parts:
\be
R^{(2),p}\,=\,C_F^2 R^{(2),p}_A \,+\, C_A C_F R^{(2),p}_{NA} \,+\, C_F T n_l R^{(2),p}_l\, +\, C_F T R^{(2),p}_F
 +\, C_F T R^{(2),p}_S
 \label{r2pdec}
\ee
where $C_A=3$, $C_F=4/3$, $T=1/2$ are group constants and $n_l=3$ is the number of light quarks.
The function $R^{(2),p}_l$ comes from the diagram with massless quark loop. It 
was found in \cite{HT} and in our normalization takes the form:
\be
R^{(2),p}_l\,=\,\left( \,-\,{1\over 4}\,\ln{\mu^2\over 4s}\, -\, {5\over 12}\, \right) R^{(1),p}\,+\,\delta^{(2)}_P
\ee
where the function $\delta^{(2)}_P$ is given by equation (110) in ref \cite{HT}.
 The function $R^{(2),p}_F$ comes from the diagram with 2 massive quark loops.  
For $s<16m^2$ it contains only the contribution
 of virtual massive quarks and has the form \cite{HT}:
\be
R^{(2),p}_F\,=\,2v\,{\rm Re}\,P^{(2)}_Q\,-\,{1\over 4}\,R^{(1),p}\,\ln{\mu^2\over m^2} \; , \qquad s<16m^2
\ee
where $P^{(2)}_Q$ is second order correction to the pseudoscalar current vertex from
the diagram with massive quark loop; it is given by equation (169) in ref \cite{HT}.
For $s>16m^2$ the 4-particle cut must be included in $R^{(2),p}_F$. It is given by the double integral,
eq.~(97) in ref \cite{HT}, which cannot be taken analytically. Here, however, the
total function $R^{(2),p}_F$ can be replaced by its high-energy asymptotic, available to
the terms $m^8/s^4$ in \cite{HS}.

The functions $R^{(2),p}_A$ and $R^{(2),p}_{NA}$ correspond to the diagrams with single 
massive quark loop and various gluon exchanges. They are not known analytically,
so one has to use some approximations. It turns out, that the moments,
computed by the dispersion relation (\ref{mompdef}), are sensitive to the choice of these
approximations. The accuracy of the moments becomes especially important in $\ov{\rm MS}$
scheme, where there is a sufficient cancellation between large terms
(see eq.~(\ref{mommsb}) below), so we describe this point in details.

The first 8 moments $M_2$--$M_9$ at $Q^2=0$ are known analytically \cite{CKS2}.
We will require, that the approximations for $R^{(2),p}_A$ and $R^{(2),p}_{NA}$
{\it must reproduce these moments} with high accuracy being substituted into the dispersion
integral in (\ref{mompdef})\footnote{We found the approximations proposed in ref \cite{CKS2},
eqs (39), (40) having insufficient accuracy to satisfy these requirement.}. As usual, we shall apply the
conformal mapping and Pade approximation for the relevant parts of the polarization function
$\Pi^p$ and take the imaginary part after then, see Appendix A for details.
Although such approximations are constructed so
that they reproduce low-$q^2$ expansion of the polarization function, they do not give exact values of
the first 8 moments at $Q^2=0$, computed by taking the dispersion integral in (\ref{mompdef}).
Indeed, the Pade approximations have extra poles away from the cut $z=[1,\infty )$ and,
strictly speaking, the dispersion relation (\ref{drpsc}) is not valid for them.
The approximated formulas for $R^{(2),p}_A$ and $R^{(2),p}_{NA}$, used in this paper,
are given in the equation (\ref{rnaapa}) of Appendix A.

The last term in (\ref{r2pdec}) $R^{(2),p}_S$ is the so-called singlet part with 2 triangle
quark loops. This part contains the 2-gluon cut, which is proportional to the 2-photon
decay width of the pseudoscalar boson. It is known analytically \cite{EGN, DSZ}:
\be
\label{ggcut}
R_{gg}^{(2),p}\,=\,TC_F {|f(z)|^2\over 2z} \,
\ee
where
$$
f(z)=-{1\over 2}\int_0^1{dx\over x} \ln{\left[1-4zx(1-x)\right]}=
\left\{\ba{cc} \arcsin^2\!\sqrt{z} & , \; z<1 \\
-{1\over 4}\left( \ln{1+v\over 1-v} -i\pi \right)^2 & , \; z>1 \ea \right.
$$
The contribution of purely gluonic states to the heavy quark current correlators 
was discussed in details in \cite{PR} (3-gluon state in case of vector currents). 
In the narrow width approximation (\ref{impop}) only charmed states are taking into account.
Since the 2-gluon state is not associated with any charmonium state,
we subtract $R_{gg}^{(2),p}$ in the dispersion relation (\ref{drpsc}) and
take the integral from $s=4m^2$, in accordance with suggestion of \cite{PR}.
The approximation for $R^{(2),p}_S$ without the 2-gluon cut
is given in the eq (\ref{rnaapa}) of Appendix A.

The leading in $\alpha_s$ order operator series $\Pi^p_{\rm OPE}$ (\ref{ppptope}) 
for the heavy quark correlator has been computed
in \cite{NR} up to operators of dimension $D=8$. This series can be compactly expressed in terms
of Gauss hypergeometric functions:
\be
\label{ppope}
\Pi^p_{\rm OPE}(s)\,=\, {1\over 2\pi^2} \sum_{k\ge 2}\sum_j
{ O_k^{(j)} \over (4m^2)^{k-2} }\sum_{i=1}^{2k-1}
    {c_{k,\,i}^{(p,\,j)} \over k! \, (1/2)_{k-1}} \,
 {}_2\!F_1\left(\left. {1, \, k+i-2\atop k-1/2} \right| {s\over 4m^2} \right)
\ee
where $(1/2)_{k-1}$ is Pochhammer symbol, $O_k^{(j)}$ are the operators of dimension $D=2k$.
For the heavy quark correlators there is single operator of dimension $D=4$
\be
O_2\,=\,\left<g^2 G_{\mu\nu}^a G_{\mu\nu}^a \right>
\ee
and 2 operators of dimension $D=6$:
\be
O_3^{(1)}\,=\,\left<g^3 f^{abc} G_{\mu\nu}^a G_{\nu\lambda}^b G_{\lambda\mu}^c \right> \; , \qquad
O_3^{(2)}\,=\,\left<g^4 j_\mu^a j_\mu^a \right>
\label{o3cond}
\ee
where $gj_\mu^a=G_{\mu\nu ; \,\nu}^a={g\over 2} \sum_{q=u,d,s} {\bar q}\gamma_\mu\lambda^a q$.
We choose 7 independent operators of dimension $D=8$ according to \cite{NR}:
\bea
O_4^{(1)} \!& = &\! \left< \left( g^2 d^{abc} G_{\mu\nu}^b G_{\alpha\beta}^c \right)^2 + \,{2\over 3}
\left( g^2 G_{\mu\nu}^a G_{\alpha\beta}^a \right)^2 \right> \; , \qquad
O_4^{(2)} \,  = \, \left< \left( g^2 f^{abc} G_{\mu\nu}^b G_{\alpha\beta}^c \right)^2\right>  \; , \nonumber \\
O_4^{(3)} \!& = &\! \left< \left( g^2 d^{abc} G_{\mu\alpha}^b G_{\alpha\nu}^c \right)^2 + \,{2\over 3}
\left( g^2 G_{\mu\alpha}^a G_{\alpha\nu}^a \right)^2 \right>\; , \qquad
O_4^{(4)} \, = \, \left< \left( g^2 f^{abc} G_{\mu\alpha}^b G_{\alpha\nu}^c \right)^2\right> \; , \nonumber \\
O_4^{(5)} \!& = &\! \left< g^5 f^{abc} G_{\mu\nu}^a j_\mu^b j_\nu^c \right> \; , \quad
O_4^{(6)} \,  = \,  \left< g^3 f^{abc} G_{\mu\nu}^a G_{\nu\lambda}^b G_{\lambda\mu; \, \alpha\alpha}^c\right>
  \; , \quad
O_4^{(7)} \,  = \, \left< g^4 j_\mu^a j_{\mu; \, \alpha\alpha}^a \right> \; .
\label{o4cond}
\eea
The coefficients $c_{k,\,i}^{(p,\,j)}$ in (\ref{ppope})  can be obtained from \cite{NR}:
$$
c_{2,\,i}^{(p)}=\left( 0,\, 1/4,\, -1/12 \right) \; , \qquad
c_{3,\,i}^{(p,\,j)}=\,
\ba{|c|ccccc|}\hline
 j\backslash i & 1 & 2 & 3 & 4 & 5\\ \hline
 1 & 0 &  -3/2 &  5 &  -22/5 &  1 \\
 2 & 1/3 &  -8/3 &  3 &  28/15 &  -4/3 \\
\hline \ea \; ,
$$
\be
c_{4,\,i}^{(p,\,j)}=\,
\ba{|c|ccccccc|}\hline
 j\backslash i & 1 & 2 & 3 & 4 & 5 & 6 & 7\\ \hline
 1 & 1/3& 31& -258& 1580/3& -365& 70& 0 \\
 2 & 1/6& 10& -111& 946/3& -4873/14& 144& -10 \\
 3 & -4/3& -16& 168& -596/3& -106& 140& 0 \\
 4 & 1/3& 122& -946& 5480/3& -7435/7& -20& 92 \\
 5 & -10/3& 12& 260& -2888/3& 7290/7& -304& -24 \\
 6 & 0& -22& 152& -174& -1108/7& 296& -96 \\
 7 & 2& -36& 100& -128& 1314/7& -208& 72 \\
\hline \ea \; .
\ee

The $\alpha_s$ correction to the $D=4$ condensate contribution
 was obtained analytically in \cite{BBIFTS}. One could differentiate
it $n$ times to obtain the moments. However, we prefer to use
a dispersion-like relation for this correction, constructed in Appendix B,
which is convenient for numerical calculation of the moments, especially for high $n$.


\section{Moments in {\boldmath $\ov{\rm MS}$} scheme}

At first let us consider the moments in the pole-mass scheme.
 In QCD the moments (\ref{mompdef}) are expanded by the running QCD coupling $\alpha_s(\mu^2)$.
 The approximation, used in this paper, includes the following ingredients:
 1) the perturbative series up to $\alpha_s^2$ order, 2) the operator series up to dimension $D=8$,
 and 3) $\alpha_s$ correction to the $\left< G^2\right>$ operator contribution. Adding all pieces
 together, we write down the following expression for the pseudoscalar moments:
\be
M_n^p(Q^2) \, = \,   \sum_{k=0}^2 M_n^{(k),p}(Q^2)\, a^k(m^2) \, + \,
 \sum_{k=2}^4 \sum_j O_k^{(j)} \,M_n^{(Ok,j),p}(Q^2)\, +\, O_2\, M_n^{(O2)(1),p}(Q^2) \, a(m^2)\; ,
\label{mpex}
\ee
for definiteness the coupling $a\equiv \alpha_s/\pi$ is taken at the scale $\mu^2=m^2$.
As discussed in previous section, the perturbative moments are taken without 
2-gluon cut (\ref{ggcut}):
\be
M_n^{(k),p}(Q^2)\,=\,4m^2 \int_{4m^2}^\infty {s\, (R^{(k),p}-R_{gg}^{(k),p})(s,m^2)\over (s+Q^2)^{n+1}}\,ds
\label{pertmom}
\ee
The leading order can be expressed in terms of Gauss hypergeometric function:
\be
 M_n^{(0),p}(Q^2) \, = \, {1\over (4m^2)^{n-2}} \,{\Gamma(n-1)\over 2\, (1/2)_n}
 \,{}_2\!F_1\!\left( \left.{n-1,\,n+1\atop n+1/2}\right| -\,{Q^2\over 4m^2}  \right)
\ee
where $(a)_n\equiv \Gamma(a+n)/\Gamma(a)$ is Pochhammer symbol.
The higher order perturbative moments are computed numerically by (\ref{pertmom}).

The contribution of the operators $O_k^{(j)}$ to the  moments can be easily obtained
by differentiating eq (\ref{ppope}):
\be
M_n^{(Ok,j),p}(Q^2)= {4\over 3} \sum_{i=1}^{2k-1}
{ c_{k,\,i}^{(p,\,j)} \over (4m^2)^{n+k-2} } \,{(k+i-2)_n \over k!\, (1/2)_{n+k-1}} \,
 {}_2\!F_1\left(\left. {n+1, \, n+k+i-2\atop n+k-1/2} \right| -{Q^2\over 4m^2} \right)
\ee
The $\alpha_s$-correction to the $D=4$ gluon condensate contribution can be obtained 
by differentiating eq (\ref{cf1pdr}) of Appendix B:
\be
M^{(O2)(1),p}_n(Q^2) \, = \, {2\over 3(4m^2)^n}\,\sum_{i=1}^3\,
{ (n+1)_{i-1}\over (i-1)!} \left[
 \, {\pi^2 f_i^p \over (1+y)^{n+i}} \, + \, \int_1^\infty  {F_i^p(z)\over (z+y)^{n+i}}\,dz\, \right]
 \label{mc1pc}
\ee
where $y=Q^2/(4m^2)$, the constants $f_i^p$ and
the functions $F_i^p(z)$ are given in eqs (\ref{fipc}) and (\ref{imfip})  of Appendix B.
Notice, that eqs (\ref{pertmom})--(\ref{mc1pc}) are applicable for noninteger $n$ also.

Similarly to the vector case \cite{IZ}, the $\alpha_s$-corrections to the moments
are unacceptably large in the pole mass scheme and the series (\ref{mpex}) is divergent.
The pole mass, in fact, is the mass of free quark. Since the quarks exist only in form of strongly
bounded states, the physical meaning of the pole quark mass is rather unclear;
it cannot be found from the sum rules with a good accuracy.

Instead of the pole mass one introduces another effective mass parameter,
to improve the convergence of the perturbative series. Authors of \cite{SVVZ,RRY}
used the mass, renormalized at the euclidean point $p^2=-m^2$.
In this paper we shall use the most popular choice for today: the gauge invariant
mass in the modified minimal subtraction ($\ov{\rm MS}$) scheme taken at the scale,
equal to the mass itself ${\bar m}\equiv {\bar m}({\bar m}^2)$.
The pole mass $m$ is perturbatively expressed in terms of ${\bar m}$:
\be
{m^2\over {\bar m}^2} \,=\, 1\,+\,\sum_{n\ge 1} \, K_n \,a^n({\bar m}^2)
\label{msbvsp}
\ee
The 2-loop factor was found, in particular, in \cite{GBGS} while the 3-loop
factor was recently calculated in \cite{MvR}:
\bea
 K_1& =&  {8\over 3}  \nonumber \\
 K_2& =&  28.6646 - 2.0828\,n_l \, = \,  22.4162 \nonumber \\
 K_3&=&  417.039 - 56.0871\, n_l + 1.3054 \, n_l^2 \, = \, 260.526
 \label{kfacnum}
 \eea
 We put $n_l=3$ in the last column.
Now we reexpand the moments (\ref{mpex}) by the QCD coupling $a({\bar m}^2)$:
\be
M_n^p(Q^2) \, = \,   \sum_{k=0}^2 {\bar M}_n^{(k),p}(Q^2)\, a^k({\bar m}^2) \, + \,
 \sum_{k=2}^4 \sum_j O_k^{(j)} \,{\bar M}_n^{(Ok,j),p}(Q^2)\, +\, 
O_2\, {\bar M}_n^{(O2)(1),p}(Q^2) \, a({\bar m}^2)\; ,
\label{mpexb}
\ee
where ${\bar M}_n^{(k),p}(Q^2)$ are expressed in terms of $M_n^{(k),p}(Q^2)$ \cite{IZ}:
\bea
{\bar M}_n^{(0),p}(Q^2) & = & M_n^{(0),p} \nonumber \\
{\bar M}_n^{(1),p}(Q^2) & = & M_n^{(1),p}-K_1 (n-d/2)\, M_n^{(0),p}+K_1(n+1)\,Q^2M_{n+1}^{(0),p}
 \nonumber \\
{\bar M}_n^{(2),p}(Q^2) & = & M_n^{(2),p}-K_1 (n-d/2)\, M_n^{(1),p}+K_1(n+1)\,Q^2M_{n+1}^{(1),p}
\nonumber \\
 & & + \, (n-d/2) \left[ {K_1^2\over 2} (n+1-d/2) - K_2 \right]  M_n^{(0),p}  \nonumber \\
 & & + \, (n+1) \left[ K_2  -K_1^2(n+1-d/2)\right] Q^2M_{n+1}^{(0),p} \nonumber \\
 & &  + \, {K_1^2\over 2} (n+1)(n+2) \,Q^4 M_{n+2}^{(0),p} \nonumber\\
{\bar M}_n^{(Ok,j),p}(Q^2) & = & M_n^{(Ok,j),p} \nonumber \\
{\bar M}_n^{(O2)(1),p}(Q^2) & = & M_n^{(O2),p}-K_1 (n+2-d/2) \, M_n^{(G,0),p} +
  K_1(n+1)\,Q^2M_{n+1}^{(G,0),p}
\label{mommsb}
\eea
where $d=4$ is the dimension of the pseudoscalar function $\Pi^p(Q^2)$,
all $M_n^{(i),p}$ in the rhs are computed with $\ov{\rm MS}$ mass ${\bar m}$.
The series (\ref{mpexb}) is much better convergent than (\ref{mpex}).
The numerical values of the ratios ${\bar M}^{(1,2)}/{\bar M}^{(0)}$ and
${\bar M}^{(O2)(1)}/{\bar M}^{(O2)}$ for $Q^2/(4{\bar m}^2)=0,1,2$ and $n=2-30$
are given in the Table 1 of Appendix C. Notice, that the values of ${\bar M}^{(2)}$
 are approximate; other approximations for $R^{(2)}$ may lead to
the moments ${\bar M}^{(2)}$, which differ from the numbers of the Table 1 within $5-10\%$.

The expansion (\ref{mpexb}) goes by $a({\bar m}^2)$.
If one takes the QCD coupling at some another scale $\mu^2$, the function ${\bar M}^{(2),p}$ changes:
\be
\label{m2shift}
a({\bar m}^2) \, \to \,   a(\mu^2)   \; , \qquad
 {\bar M}_n^{(2),p}(Q^2) \, \to \, {\bar M}_n^{(2),p}(Q^2) \,+
 \,  {\bar M}_n^{(1),p}(Q^2)\, \beta_0 \, \ln{ \mu^2\over{\bar m}^2}
\ee
so that the series (\ref{mpexb}) is $\mu^2$-independent at the order $\alpha_s^2$.


\section{Pseudoscalar sum rule}

It is convenient to define a dimensionless ratio of the pseudoscalar moments:
\be
\label{rdimles}
r_n(Q^2)\,=\,{1\over 4{\bar m}^2}\,{M_n^p(Q^2)\over M_{n+1}^p (Q^2)}\,
\to\,{m_{\eta_c}^2+Q^2\over  4{\bar m}^2} \; , \qquad n\to \infty
\ee
Theoretical ratio depends on the quark mass ${\bar m}$, QCD coupling $\alpha_s$ and condensates.
But if the dimensionless parameters $Q^2/(4{\bar m}^2)$, $\left<aG^2\right>/(4{\bar m}^2)^2$
etc.~are fixed, the
l.h.s.~of (\ref{rdimles}) does not depend on the quark mass ${\bar m}$ (in fact, the QCD coupling
depends on the scale, which itself may depend on ${\bar m}$; but this dependence is weak within the
range of error of ${\bar m}$). So one may use the ratio (\ref{rdimles}) to find the $\ov{\rm MS}$ charm
quark mass ${\bar m}$ for given condensates and QCD coupling.

The QCD coupling constant $\alpha_s$ is universal value and can be taken from other experiments.
As input parameter, it is convenient to take $\alpha_s$ at the $\tau$-lepton mass \cite{PDG}:
\be
\label{altau}
\alpha_s(m_\tau^2)\,=\,0.33\pm 0.03 \; , \qquad m_\tau\,=\,1.777 \, {\rm GeV}
\ee
Using this value as the boundary condition in the renormalization group equation, the
QCD coupling can be evaluated at any scale. As argued in \cite{IZ}, the most natural scale for
$\alpha_s$ is
\be
\label{ascale}
\mu^2\,=\,Q^2\,+\,{\bar m}^2
\ee
Indeed, in the limit $Q^2\gg {\bar m}^2$ we come to natural massless choice $\alpha_s(Q^2)$, while
at $Q^2=0$ it becomes $\alpha_s({\bar m}^2)$. Later we shall vary the scale (\ref{ascale})  to check
the stability of results.

The $\ov{\rm MS}$ charm quark mass is determined from vector charmonium sum rules with high
accuracy. The analysis of the moments at $Q^2=0$ with $\alpha_s^2$ corrections leads to
${\bar m}=1.304\pm 0.027\,{\rm GeV}$ in \cite{KS} and
$1.23\pm 0.09\,{\rm GeV}$ in \cite{EJ}. The authors of \cite{KS} neglected the condensate
contribution, while in \cite{EJ} the value $\left< aG^2\right>=0.024\pm 0.012\,{\rm GeV}^4$
was employed. In fact, the gluon condensate weakly affects on the mass value. But for the
condensate determination the mass accuracy is especially important: a small mass variation
leads to significant condensate change. As noticed in \cite{IZ}, the perturbative $\alpha_s$
and $\alpha_s^2$ corrections to the vector moments $M_n(Q^2)$ in $\ov{\rm MS}$ scheme are
strongly suppressed for $Q^2/(4{\bar m}^2)\approx n/5-1$ and $n>5$. The analysis of \cite{IZ}
at $Q^2/(4{\bar m}^2)=1,2$ allowed to determine the $c$-quark mass with high acuracy:
\be
\label{mcmsb}
{\bar m}({\bar m})\,=\,1.275\pm 0.015 \; {\rm GeV}
\ee
independently of the condensate value. (If the condensate is fixed, the error in (\ref{mcmsb})
can be reduced even further.) This result is close to the recent lattice calculation
${\bar m}=1.26\pm 0.04(stat)\pm 0.12(syst)\,{\rm GeV}$ \cite{BLM}.

For the mass (\ref{mcmsb}) one gets the ratio (\ref{rdimles})
\be
\label{rex}
r_n(Q^2)\,=\,{Q^2\over 4{\bar m}^2}\,+\,1.37\pm 0.03
\ee
in the limit $n\to\infty$. Which values of $Q^2/(4{\bar m}^2)$  are convenient for the sum rule analysis?
The choice $Q^2=0$ is not appropriate, since the perturbative corrections to the moments
are large for almost all $n$, even in $\ov{\rm MS}$ scheme.
Large $Q^2 \gtrsim 12{\bar m^2}$ are also dangerous: in particular,
when one changes the scale of $\alpha_s$ in (\ref{m2shift}), the effective expansion
parameter $a\beta_0\ln{(Q^2/{\bar m}^2)}$ becomes large $\gtrsim 0.5$. In what follows
we shall use two choices $Q^2/(4{\bar m}^2)=1,2$.

\begin{figure}[tb]
\epsfig{file=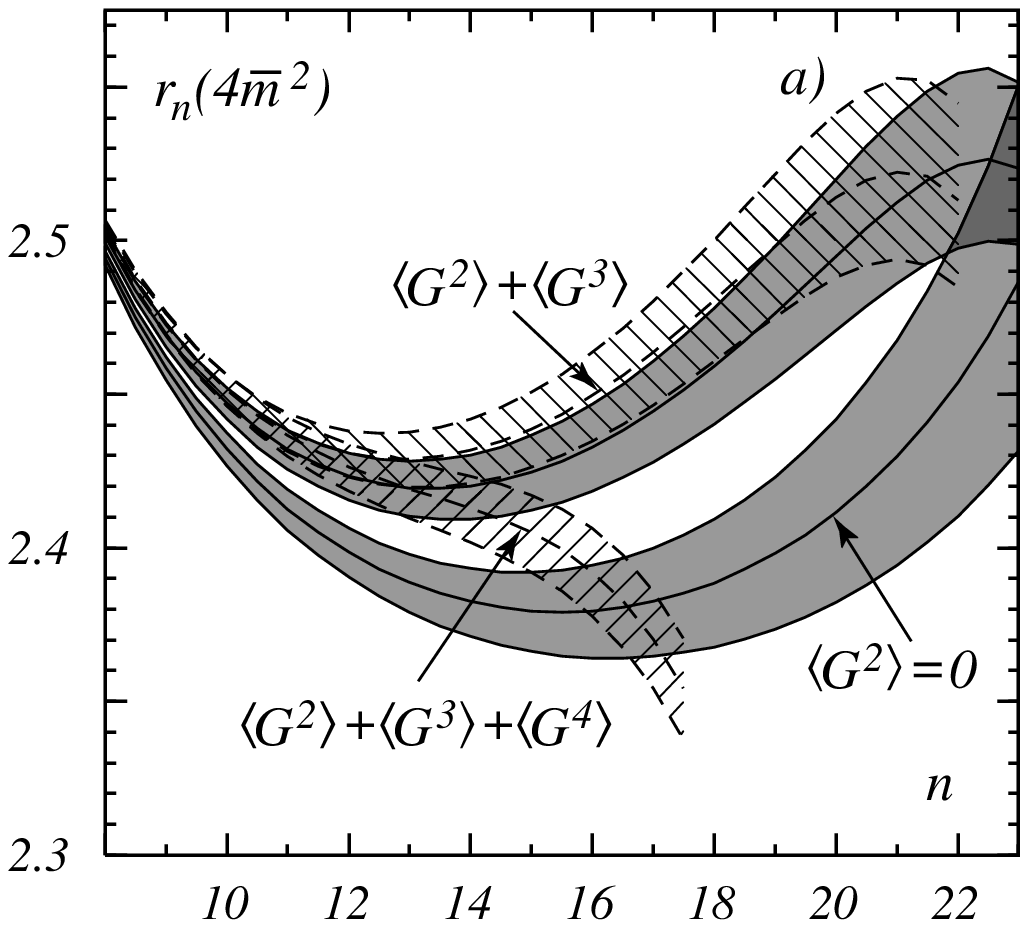, width=80mm}\hspace{4mm}\epsfig{file=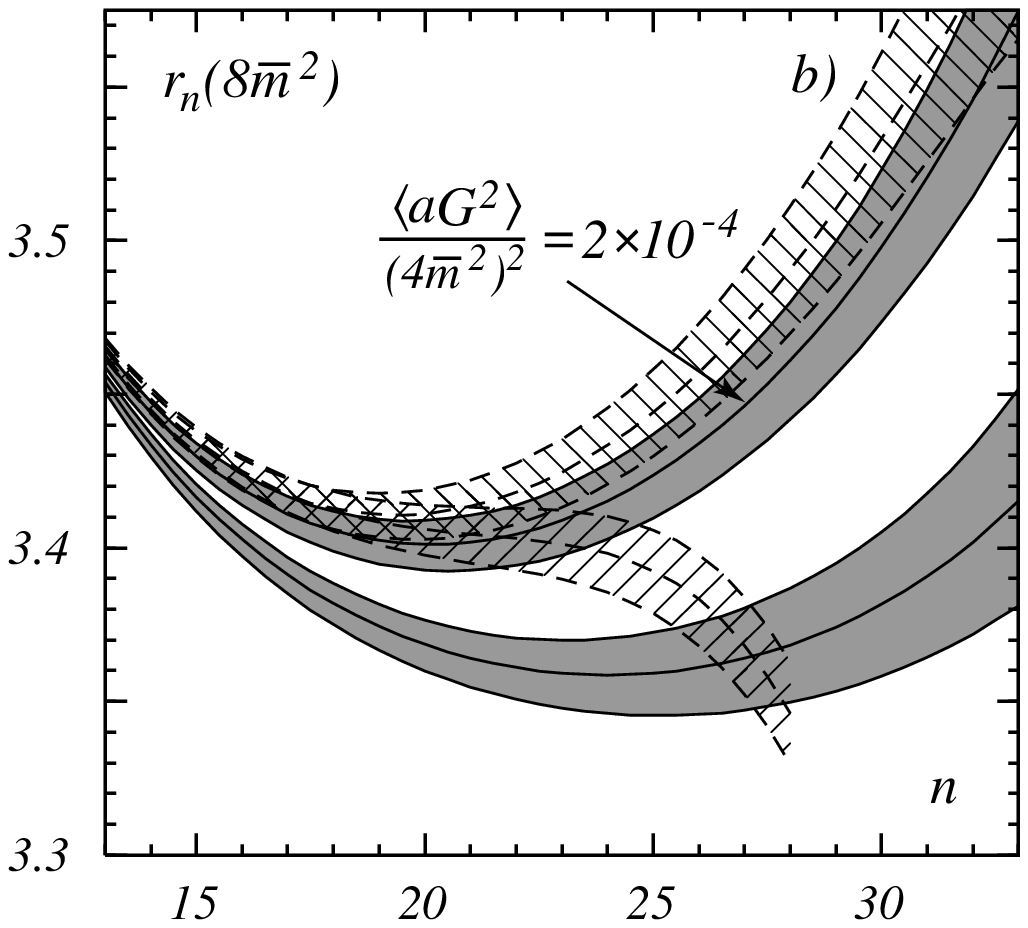, width=80mm}
\caption{Ratio $r_n(Q^2)$ for $Q^2=4{\bar m}^2$ a) and $Q^2=8{\bar m}^2$ b) versus $n$.
Lower shaded curve is purely perturbative, upper shaded curve is computed
with condensate $\left<aG^2\right>/(4{\bar m}^2)^2=2\times 10^{-4}$.
Hatched curves include the $\left<G^3\right>$ operator (upper curve)
and $\left<G^3\right>+\left<G^4\right>$ operators (lower curve) computed
in the instanton model (\ref{g4inst}). The errorband of each curve corresponds to the error of $\alpha_s$ in
(\ref{altau}).}
\label{fig_r12}
\end{figure}

The theoretical ratios $r_n(4{\bar m}^2)$ and $r_n(8{\bar m}^2)$
 are plotted versus $n$ in the Fig \ref{fig_r12}a) and \ref{fig_r12}b) respectively.
The lower shaded curve is purely perturbative, i.e.~for $\left<aG^2\right>=0$.
The central line of the shaded area corresponds to the central value of $\alpha_s$ (\ref{altau}),
the errorband covers the error of the coupling $\alpha_s$ in (\ref{altau}).
One sees, that the agreement with (\ref{rex}) is achieved within relatively narrow range of $n$:
$n\sim 16$ for $Q^2=4{\bar m}^2$ and $n\sim 24$ for $Q^2=8{\bar m}^2$.
If we look at the Table 1, the perturbative corrections to the moments in $\ov{\rm MS}$ scheme,
as well as $\alpha_s$ correction to the condensate contribution, are minimal here.
For higher $n$ these corrections grow rapidly and the perturbation theory cannot be trusted here.
For lower $n$ the perturbative corrections are also large, and the leading order of the $D=4$
condensate contribution crosses 0 at some point, so the behavior of the $\alpha_s$-series
is rather unclear here. Moreover, unknown contribution of $\eta_c(2S)$ and higher states to the
experimental moments could be significant for low $n$.

Now we consider nonzero $D=4$ condensate. As an illustration, let us fix the ratio
$\left<aG^2\right>/(4{\bar m}^2)^2$ $=2\times 10^{-4}$, which corresponds to
$\left<aG^2\right>\approx 0.008\, {\rm GeV}^4$,
close to the central value obtained in \cite{IZ}. The ratio $r_n$ with this condensate
is shown by the upper shaded curves in Fig \ref{fig_r12}.
The ratio becomes higher for nonzero condensate, which tells in favor of lower mass of $c$-quark.
At $Q^2=4{\bar m}^2$ the ratio is even higher, than (\ref{rex}) for all $n$.

Several models were employed to estimate the higher dimension $D=6,8$ operators.
Here we consider the dilute instanton gas model \cite{NSVZ} and vacuum dominance (or factorization)
model.

{\bf Instanton model.} The vacuum configuration is codsidered as a dilute gas of noniteracting
instantons with effective radis $\rho_c$ and concentration $n_0$. The radius $\rho_c$
varies within $0.3-1\,{\rm fm}$ in the literature. Here we shall use the estimation
$\rho_c=0.5\,{\rm fm}$ obtained in \cite{IS}. The instanton concentration $n_0$ is fitted to
the $D=4$ gluon condensate $\left<aG^2\right>=32\,n_0$. Then one obtains the following expressions
for the $D=6,8$ gluon condensates (\ref{o3cond}), (\ref{o4cond}):
$$
O^{(1)}_3\,=\,{12\over 5\rho_c^2}\,O_2 \; , \qquad O^{(2)}_3\,=\, 0 \; ,
$$
\be
\label{g4inst}
\left( O_4^{(1)}, \ldots , O_4^{(7)} \right) = \left( 4,8,3,4,0,8,0 \right)
{16\over 7\rho_c^4} \,O_2
\ee
The ratio $r_n$ with the operators $O_{3,4}$ computed by the instanton model (\ref{g4inst})
is show by hatched curves in Fig \ref{fig_r12}. Upper hatched curve includes $\left<G^3\right>$
operator only, lower hatched curve includes both $\left<G^3\right>$  and $\left<G^4\right>$.

The $\left<G^3\right>$  operator contribution is small. But the contribution of the $\left<G^4\right>$
operators is large in the region of interest. Obviously the place, where the lower hatched curve
crosses the perturbative one (lower shaded), the sum rule (\ref{rdimles}) with the operators (\ref{g4inst})
is not applicable. Here the $\left<G^4\right>$ contribution exceeds the leading order  $\left<G^2\right>$,
and the OPE series diverges.

It is a demonstration, that the higher order operators are essentialy overestimated in the instanton 
model \cite{IS}. Moreover, their values strongly depend on the instanton size $\rho_c$, which is 
not strictly fixed. For this reason we finish the analysis in the instanton model. The main outcome 
of this analysis is relatively small contribution of the operator $\left<G^3\right>$, which will be 
ignored in what follows.

{\bf Factorization hypothesis}. In the factorization model the $\left<G^4\right>$
operators are proportional to $\left(\left<G^2\right>\right)^2$. The operators with the light quark current
$j_\mu^a$ in (\ref{o3cond}), (\ref{o4cond}) can also be estimated by the factorization, but their size
is much smaller. The operator $O_4^{(6)}$ with derivatives was taken as $O_4^{(6)}\approx M^2 O_3^{(1)}$
in \cite{NR}, where $M^2 \approx 0.3 \, {\rm GeV}^2$ characterizes the gluon virtuality in the vacuum.
Alternatively, one may express this operator as
$$
O_4^{(6)}\, =\, 2\left< g^4 f^{abc} G^a_{\mu\nu} G^b_{\nu\lambda} j^c_{\lambda ; \mu}\right> +\, 2\, O_4^{(4)}
$$
Since we neglect the operators with $j_\mu^a$, we take $O_4^{(6)}=2O_4^{(4)}$ here
(both estimations agree in the order of magnitude for typical condensates). Summarizing,
we write down the $D=8$ operators as:
\be
\label{o4fac}
\left( O_4^{(1)}, \ldots , O_4^{(7)} \right) =
\left(\, {65\over 144}, \, {5\over 16}, \, {19\over 72}, \, {1\over 16},\, 0,\, {1\over 8},\, 0\, \right) \left(O_2\right)^2
\ee
The accuracy of the factorization is expected to be $\sim 1/N_c^2$, $N_c=3$ is the color number.
(The $1/N_c^2$ ambiguity of the $D=8$ quark-gluon condensate factorization was explicitly
demonstrated in \cite{IZ2}.) Another version of the factorization, which employs the heavy quark expansion,
was proposed in \cite{BLPT}.

\begin{figure}[tb]
\epsfig{file=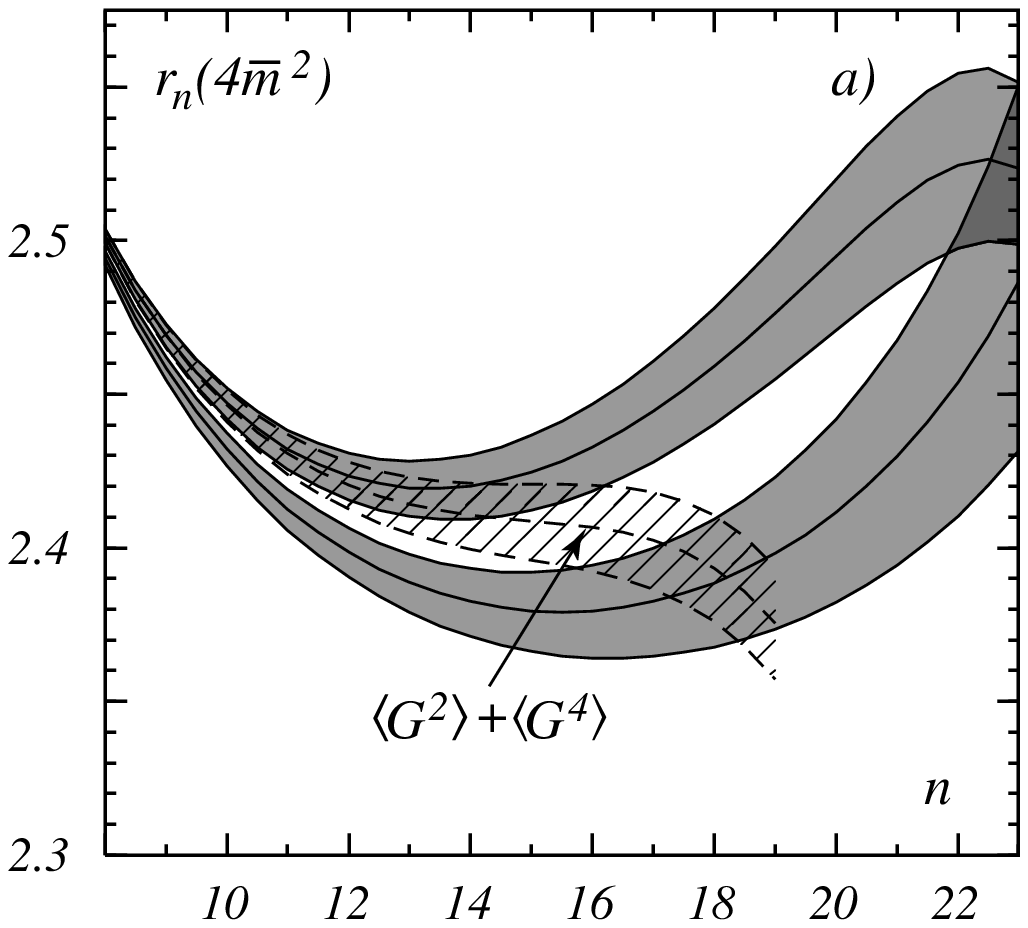, width=80mm}\hspace{4mm}\epsfig{file=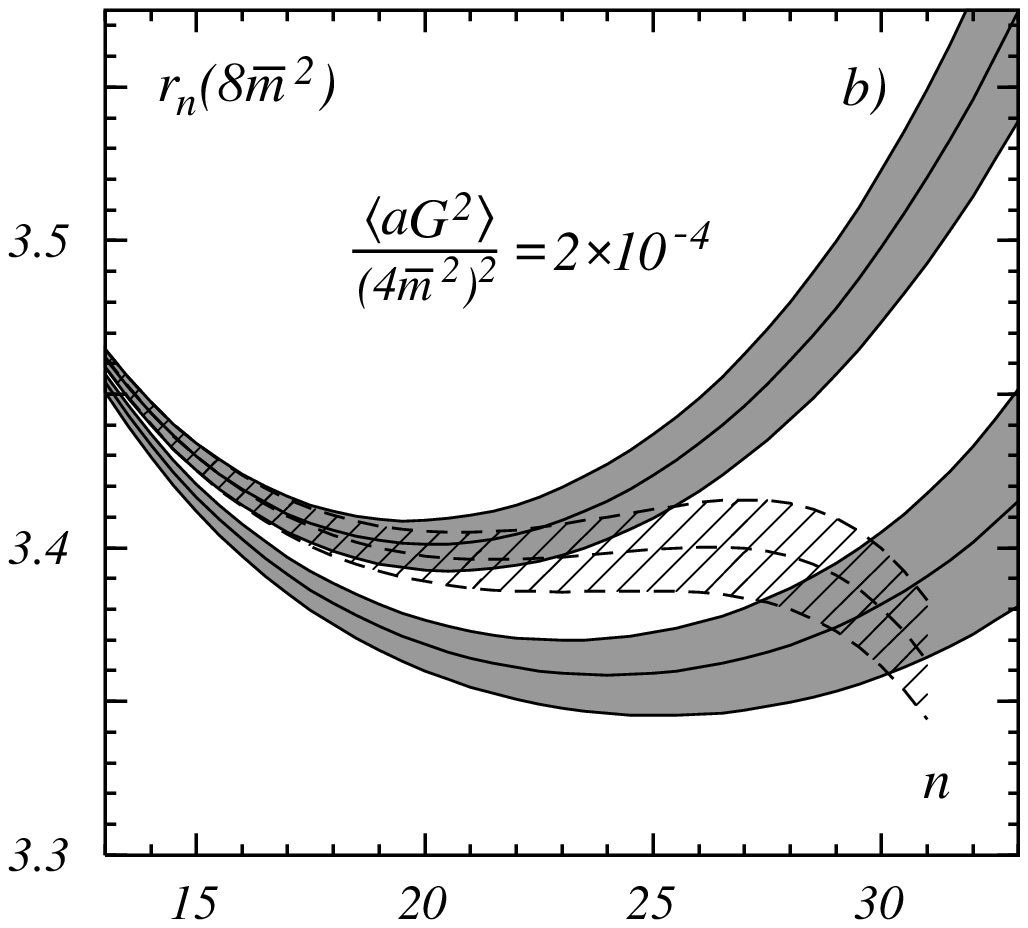, width=80mm}
\caption{Ratio $r_n(Q^2)$ for $Q^2=4{\bar m}^2$ a) and $Q^2=8{\bar m}^2$ b) versus $n$ in the 
factorization model. The $\left< G^3\right>$ condensate is neglected, the contribution of the 
$\left< G^4\right>$ operators according to (\ref{o4fac}) is displayed by the hatched curve. Other 
notations are the same as in Fig \ref{fig_r12}.}
\label{fig_r12f}
\end{figure}

The ratio $r_n$ with the operators (\ref{o4fac}) is shown by the hatched curves in the Fig \ref{fig_r12f}
for the $D=4$ gluon condensate $\left< G^2 \right>/(4{\bar m}^2)^2=2\times 10^{-4}$. 
Comparing the Figures \ref{fig_r12} and \ref{fig_r12f} one sees, 
that the contribution of the $\left< G^4 \right>$ operators in the factorization model
is smaller than in the instanton one. It allows to establish certain stability region, 
where the ratio $r_n$ remains almost unchanged. This region is clearly visible for $Q^2=8{\bar m}^2$:
at $n=20-26$  the ratio is $r_n=3.40\pm 0.01$, which corresponds to the $c$-quark mass
${\bar m}=1.260\pm 0.005 \, {\rm GeV}$.  This mass is computed for the condensate 
$\left< aG^2 \right>=0.008\,{\rm GeV}^4$. 

In the same way the mass can be computed for other values of the condensate. A restriction on the
charm quark mass for different condensates is calculated in the next section.


\section{Restrictions on the {\boldmath $c$}-quark mass and {\boldmath $D=4$} gluon condensate}

As the main result of the pseudoscalar charmonium sum  rule (\ref{raps0}), we may establish
certain restrictions for the $c$-quark mass ${\bar m}$ for a given condensate $\left<aG^2\right>$. 

At first, let us neglect the higher dimension operators $\left< G^3 \right>$ and  $\left< G^4 \right>$. 
The calculation goes as follows. For a given $Q^2/(4{\bar m}^2)$ one should
establish the range of $n$, where the perturbation theory as well as operator expansion can be trusted.
It is reasonable to require, that the perturbative corrections may not exceed $30-40\%$
of the leading term. The most dangerous is
the $\alpha_s$-correction to the gluon condensate contribution ${\bar M}^{(G,1)}_n$.
Keeping in mind typical size of the QCD couping $\alpha_s/\pi\sim 0.1$,
let us impose the restriction $|{\bar M}^{(G,1)}_n/{\bar M}^{(G,0)}_n|<4$.
From the Table 1 we find the following range of $n$:
\be
\label{regn}
n=14-19 \; \; \;  {\rm for} \; \; \; {Q^2\over 4{\bar m}^2}=1 \qquad {\rm and} \qquad
n=22-30 \; \; \; {\rm for} \; \; \; {Q^2\over 4{\bar m}^2}=2.
\ee
The perturbative corrections to the moments
are also tolerable in this region: the first correction  $|{\bar M}^{(1)}_n/{\bar M}^{(0)}_n|<2.5$
and the NNLO correction $|{\bar M}^{(2)}_n/{\bar M}^{(0)}_n|<17$.
Then, we take some value of $\left<aG^2\right>/(4{\bar m})^2$ and find the maximal and minimal
value of the ratio $r_n(Q^2)$ within this range of $n$. From these numbers we find the
minimal and maximal values of the charm quark mass ${\bar m}$.

\begin{figure}[tb]
\hspace{2mm} \epsfig{file=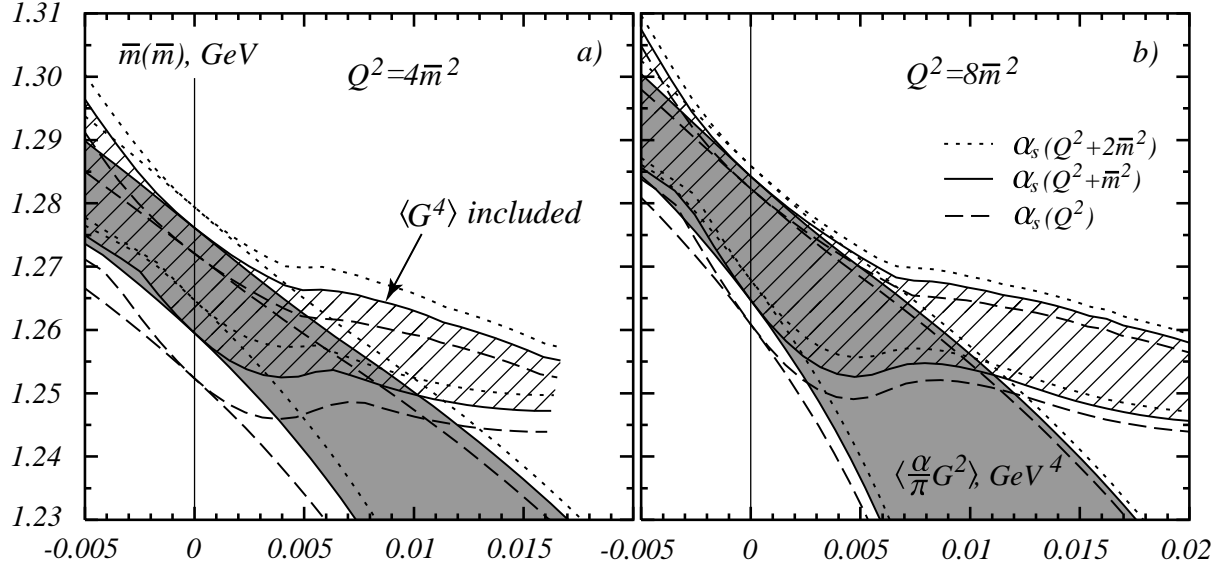, width=160mm}
\caption{Charm quark mass ${\bar m}({\bar m})$ versus $\left<aG^2\right>$ obtained from
the pseudoscalar sum rule. Shaded area displays the acceptable region when the $\left< G^4\right>$
condensates are neglected. Hatched  region corresponds to the $\left< G^4\right>$ condensates,
computed by factorization model. Dashed lines show the region boundaries for 2 alternative choices of 
the $\alpha_s$ scale.}
\label{fig_mvsg}
\end{figure}

The results are shown by the shaded regions in the Fig \ref{fig_mvsg}.
Fig \ref{fig_mvsg}a) and \ref{fig_mvsg}b) display the restrictions, obtained from the sum rule (\ref{rdimles})
at $Q^2=4{\bar m}^2$ and $Q^2=8{\bar m}^2$ respectively.
Since unknown higher order in $\alpha_s$ moments are discarded everywhere,
the results depend on the choice of the scale, at which $\alpha_s$ is taken.
The dark area shows the acceptable region for the scale (\ref{ascale}).
The dashed and dotted lines display the boundaries of the acceptable region, if
${\bar m}^2$ is added to or subtracted from this scale. The scale dependence is weaker at 
$Q^2=8{\bar m}^2$.

It is clear from the Fig \ref{fig_mvsg}, that the pseudoscalar sum rule prefers lower values 
of the gluon condensate. In particular, for the mass ${\bar m}=1.275\pm 0.015\,{\rm GeV}$ 
\cite{IZ} one obtains the upper condensate limit $\left< aG^2 \right> <0.008 \, {\rm GeV}^4$.

Now let us include the higher dimension operators. As follows from the instanton model analisys, the
contribution of the $D=6$ condensate $\left< G^3 \right>$ is small in the region of interest. But the
$D=8$ operators change the ratio $r_n$ essentially. At some $n$ their contribution exceeds the
leading condensate $\left< G^2 \right>$; at this point the OPE series is divergent. 
Let us require that the contribution of the $D=8$ operators $O_4^{(i)}$ to the moments may 
not exceed $30\%$ of the $D=4$ condensate contibution. This requirement further reduces
the region of $n$ (\ref{regn}) depending on the condensate size. In the factorization model 
$\left< G^4 \right> \sim \left( \left< G^2 \right> \right)^2$, and the region becomes smaller for
higher condensate $\left< G^2 \right>$.

The hatched regions in the Fig \ref{fig_mvsg} display the inclusion of the $\left< G^4 \right>$ 
operators in the factorization model. For $\left< aG^2 \right> >0.005\, {\rm GeV}^4$ the $D=8$ 
operators  change the ratio $r_n$ drastically. They compensate the leading condensate 
and the ratio $r_n$ becomes almost independent of the $\left< aG^2 \right>$ condensate
in the stability region. From the hatched area in the Fig \ref{fig_mvsg} one gets the following 
limits of the $c$-quark mass:
\be
\label{mc}
{\bar m}_c({\bar m}_c)\,=\,1.26\pm 0.02 \, {\rm GeV}
\ee
indenpendently on the condensate value. The mass (\ref{mc}) is in agreement with the result
${\bar m}=1.275\pm 0.015\,{\rm GeV}$, obtained from the vector charmonium sum rules 
in \cite{IZ}.

If the $\left< G^4 \right>$ operators are included,
it becomes rather difficult to obtain certain restrictions on the condensate value.
However, for large condensate the stability region is narrow, and the
results become unreliable. In particular, for $\left< aG^2 \right> > 0.015\,{\rm GeV}^4$
and  $Q^2=4{\bar m}^2$ there is no region of $n$, where the OPE series looks convergent.
This sets the natural limit of the condensate value, at which the pseudoscalar sum rule 
works. 

\section*{Acknowledgement}

Author thanks B.L.Ioffe for discussions. The research described in this publication
was made possible in part by Award No RP2-2247
of U.S. Civilian Research and Development Foundation for Independent States of Former
Soviet Union (CRDF), by the Russian Found of Basic Research grant 00-02-17808 and
INTAS grant 00-00334.


\section*{Appendix A: Approximations for {\boldmath $R^{(2),p}$}}

Let us  define the dimensionless coefficient functions for the
perturbative correlator (\ref{ppptope}) as follows:
\be
\Pi^p(q^2)=4m^2q^2\sum_{k\ge 0} \Pi^{(k),p}(q^2) a(m^2)  \; , \qquad
\Pi^{(k),p}(q^2)={3q^2\over 8\pi^2}\int_0^\infty {R^{(k),p}(s,m^2)\over s (s-q^2)}ds \; ,
\ee
for definiteness we take the QCD coupling at the scale $\mu^2=m^2$ and put the constants
$c_0=c_1=0$ in the dispersion relation (\ref{drpsc}). The 3-loop function $\Pi^{(2),p}$ is decomposed
into 5 gauge invariant parts in the same way as $R^{(2),p}$ (\ref{r2pdec}).

At first we consider the nonabelian part $\Pi^{(2),p}_{NA}$. Its expansion near $z\equiv q^2/(4m^2)=0$
until $z^8$ is available in \cite{CKS2}. Then, as usual, we reexpand this series in terms of the
variable $\omega$, which naturally appears in the perturbative calculations:
\be
\omega\,=\,{1-\sqrt{1-z}\over 1+\sqrt{1-z}}
\ee
The expansion of the polarization operator in $\omega$ has
appropriate analytical properties, namely the cut at $z=[1,\infty)$.
In many cases the Pade approximation was proved to have better accuracy, than Tailor series.
The best results (see the discussion in Section 2) were obtained for the Pade approximation [5/2]:
$$
\Pi^{(2),p}_{NA}(\omega)\,  =\, {3\over 16\pi^2} \times \hspace{110mm}
$$
\be\label{pinapa}
\times {-7.43220\omega + 73.5001\omega^2 +
   2.69248\omega^3 -13.3868\omega^4 - 0.91032\omega^5 -
   1.61120\omega^6 \over  1 - 0.18579\omega - 0.63849\omega^2 }
\ee

The accuracy of the Pade approximated abelian part $\Pi^{(2),p}_A$ is worse because
of Coulomb behavior $\sim 1/\sqrt{1-z}$ near the threshold. It turns out, however,
that the expansion in $\omega$ converges faster, if the multiplier $1/(1-\omega^2)$
is separated out:
\bea
\Pi^{(2),p}_A(\omega)&  = & {3\over 16\pi^2}\left[
10.8487 \,\omega + 90.9348 \,\omega^2 + 111.510 \,\omega^3 +
 73.3467 \,\omega^4  \right. \nonumber \\
 & & \hspace{8mm} \left.+ 113.363 \,\omega^5 + 74.8959 \,\omega^6 + 114.366 \,\omega^7
 + 77.2105 \,\omega^8 + O(\omega^9)\,\right] \nonumber \\
 &  = & {3\over 16\pi^2}{1\over 1-\omega^2}\left[
 10.8487 \,\omega + 90.9348 \,\omega^2 + 100.662 \,\omega^3 -
 17.5881 \,\omega^4 \right. \nonumber \\
 & & \hspace{10mm}\left .+ 1.85289 \,\omega^5 + 1.54919 \,\omega^6 +
 1.00228 \,\omega^7 + 2.31461 \,\omega^8  + O(\omega^9) \, \right] \nonumber
 \eea
 Now we construct the Pade approximation, which well reproduces all asymptotic
 and first 8 moments from the dispersion relation:
$$
\Pi^{(2),p}_A(\omega)\,  =\, {3\over 16\pi^2} \times \hspace{110mm}
$$
\be \label{piapa}\times
 {10.8487\,\omega + 145.611\,\omega^2 +
   507.376\,\omega^3 + 57.361\,\omega^4 - 565.406\,\omega^5 +
   94.514\,\omega^6 \over  (1 - \omega^2)\, (1 + 5.03984\,\omega - 4.75471\,\omega^2) }
\ee

 Few more work should be done to construct the approximation for the singlet
 polarization function $\Pi^{(2),p}_S$. Its expansion near $z=0$ until $z^8$ is available in \cite{CHS}.
 The singlet correlator contains intermediate massless 2-gluon state, so the
 cut starts from $z=0$, the expansion in \cite{CHS} has the terms $\sim\ln{(-z)}$
  and the conformal mapping procedure ($z\to \omega$) is not applicable here.
 As discussed in Section 2, in our sum rules we use the polarization operator 
 {\it without 2-gluon cut} (\ref{pertmom}),
so the correspondent part of the polarization operator should be subtracted from the result of
\cite{CHS}:
\be
\Pi^{(2),p}_{gg}\,=\,{3\over 16\pi^2} C_F T (-y) \int_0^\infty {|f(z)|^2\over z^2(z+y)}dz \; , \qquad
y={Q^2\over 4m^2}
\ee
where the function $f(z)$ is given in (\ref{ggcut}). The integral from $z=1$ to $\infty$ is regular at
$y=0$ and can be expanded by $y$ in Tailor series. But the integral from $z=0$ to 1 requires special
care, since it behaves as $\sim\ln{y}$ at $y\to 0$. In order to obtain the expansion for small $y$,
we suggest to use the following series for the function $|f(z)|^2$ for $0<z<1$ \cite{PBM}:
\be
{\rm arcsin}^4\!\sqrt{z}\,=\,{3\over 2}\sum_{n=2}^\infty
 \,z^n\, {(n-1)!\over n\,(1/2)_n} \,\sum_{k=1}^{n-1}{1\over k^2}
\ee
Then we obtain the following expansion:
\be
\int_0^1 { {\rm arcsin}^4\!\sqrt{z}\over z^2(z+y)}dz \, =\, -\, {{\rm arcsh}^4\!\sqrt{y}\over y^2}\,\ln{y}\,+\,
\sum_{n=0}^\infty \, (-y)^n\, I_n \; ,
\ee
where the constants
$$
I_n\,=\,{3\over 2}\sum_{k=2,\ne n+2}^\infty {(k-1)!\over k(k-n-2)\,(1/2)_k}\sum_{j=1}^{k-1}
{1\over j^2}
$$
can be computed analytically for any $n$ with the help of recursive relations:
\bea
I_n & =& -{\pi^4\over 16(n+2)} + {2\,(n+1)!\over (n+2)\,(5/2)_n} \left\{ \pi^2\ln{2} -
{7\over 2}\zeta_3 + \sum_{k=1}^{n+1} \left[ {1\over k^2(n+2)} + {2\over k^3} \right. \right.  \nonumber \\
 & & \left. \left.
 -\,{(1/2)_k\over k^2\,(k-1)!}{\pi^2-8\over 2}\,
 +\,{1\over k^2} \sum_{j=1}^{n+1}\left({1\over j+1/2} -{1\over j} \right)
 -\,{1\over k^2} \sum_{j=1}^{k-1} {(3/2)_j\over j^2\,(j-1)!}\,\right] \right\}
\eea
Now one obtains regular at $y=0$ Tailor expansion of the polarization operator $\Pi^{(2),p}_S$
without the 2-gluon cut, applies the conformal mapping and constructs the Pade approximation:
$$
\Pi^{(2),p}_{S-gg}(\omega)  \equiv  \Pi^{(2),p}_S   - {1\over C_F T} \Pi^{(2),p}_{gg}  = {3\over 16\pi^2} \times
\hspace{60mm}
 $$
 \be
 \times {-7.86155\,\omega + 6.98952 \,\omega^2 + 3.24217 \, \omega^3 -
  1.68013\, \omega^4 - 0.31547 \, \omega^5 - 0.00464\, \omega^6 \over 1 -
  0.05236 \, \omega - 0.32689 \,\omega^2} \label{pispa}
\ee

Eventually we take the imaginary part of (\ref{pinapa}), (\ref{piapa}), (\ref{pispa})
and obtain the correspondent coefficient functions $R^{(2),p}$:
\bea
R^{(2),p}_{NA}&  = & {8\pi\over 3} \, {\rm Im} \,\Pi^{(2),p}_{NA}(\omega)
  + {11\over 16}\,R^{(1),p}\ln{\mu^2\over m^2} \; ,  \nonumber \\
R^{(2),p}_A&  = & {8\pi\over 3} \, {\rm Im} \,\Pi^{(2),p}_A(\omega)  \; , \nonumber \\
R^{(2),p}_S -{1\over C_F T} R^{(2),p}_{gg}&  = & {8\pi\over 3} \, {\rm Im} \,\Pi^{(2),p}_{S-gg}(\omega)
   \; , \qquad {\rm where} \qquad  \omega={1+i\sqrt{z-1} \over 1-i\sqrt{z-1}}  \label{rnaapa}
 \eea


\section*{Appendix B: {\boldmath $\alpha_s$}-correction to the condensate contribution}

The $\alpha_s$ correction to the $D=4$ gluon condensate contribution was found
 in \cite{BBIFTS}. Let us parametrize it by dimensionless function $f^{(1),p}(z)$:
\be
\Pi^p_{\rm OPE}(s)\,=\,\ldots \,+\, {O_2\over 4\pi^2}\,a\,f^{(1),p}(z) \; ,
\ee
where dots denote the leading order operator contribution (\ref{ppope}).
Here we construct a dispersion-like relation for this function,
 convenient for numerical calculation of the moments. We will follow the method,
 used in \cite{IZ} for the vector current correlator.

 The imaginary part is:
\be
\label{drcf1p}
 {\rm Im}\, f^{(1),p}(z+i0)= {\pi\over 96 z^3 v^5}\left[ P_2^P(z) + {P_3^P(z)\over zv} \ln{1-v\over 1+v}
 +P_4^P(z)(1-z)\left(2\ln{v}+{3\over 2}\ln{(4z)}\right)\right]
\ee
where the polynomials $P_i^P(z)$ are given in the Table 1 of ref \cite{BBIFTS}.
Taking the contour integral in $z$-plane around the cut $z=[1,\infty)$, one could write down
 the following dispersion relation:
 \bea
  f^{(1),p}(t) & = & {1\over \pi}\int_{1+\eps}^\infty  {{\rm Im}\, f^{(1),p}(z+i0)\over z-t} dz \,
  + \, \sum_{i=1}^3 {\pi^2 f_i^p\over (1-t)^i}
  \,-\, {11\over 384}\,{\eps^{-3/2}\over 1-t } \nonumber \\
  & &
 +   \left[ {15473\over 6912} + {35 \over 36}\ln{(8\eps)}\right] {\eps^{-1/2}\over 1-t}
    \,+\, {11 \over 128}\, { \eps^{-1/2} \over (1-t)^2}    \label{cif1p}
\eea
   where $\eps\to 0$ and
   \be \label{fipc}
   f_1^p\,= \,-\,{11\over 256} \; , \qquad    f_2^p\,=\,{69 \over 256}\; , \qquad    f_3^p\,=\,- \, {197\over 2304} \; .
   \ee
   To simplify the calculation, we represent the imaginary part (\ref{drcf1p})   in the following form:
\be
{1\over \pi} {\rm Im}\, f^{(1),p}(z+i0) \,=\, F_1^p(z)\,+\,{F_2^p}'(z)\,+\,{1\over 2}\,{F_3^p}''(z)
\ee
where the functions $F^i_p(z)$ grow not faster than $(z-1)^{-1/2}$ at $z\to 1$
 and have appropriate asymptotic
at infinity. Our choice is:
\bea
 F_1^p(z) & = &  {1\over 96 \,z^2 v}\left[
 56 + {49\over 3}z + 552 z^2 \right.
 + \left({207\over 4} -60z + 904 z^2 - 4216 z^3 + 3312 z^4 \right) \times \nonumber \\
 & & \times {1\over zv} \ln{1-v\over 1+v}
 + \left. \left( {140\over 3} + {76\over 3} z - {5120 \over 3} z^2 + 2208 z^3\right)
 \left(2\ln{v}+{3\over 2}\ln{(4z)}\right)\right] \nonumber \\
F_2^p(z) & = &  {1\over 96 \,z^2 v}\left[
 {11\over 4} + {645\over 2}z   \right.
  + \left(-{197\over 24} + {1439 \over 24}z\right) {1\over zv} \ln{1-v\over 1+v} \nonumber \\
& & + \left. {280 \over 3} z\left(2\ln{v}+{3\over 2}\ln{(4z)}\right)\right] \nonumber \\
F_3^p(z) & = &  {1\over 96 \,z v}\left[
 -{131\over 6} - {197\over 12\, zv} \ln{1-v\over 1+v} \right]
\label{imfip}
 \eea
 Then, the r.h.s.~of (\ref{cif1p}) can be integrated by parts twice, 
 all divergent in $\eps\to 0$ terms cancel and
 the dispersion relation can be brought to the following form:
\be
f^{(1),p}(t) \, =\,\sum_{i=1}^3 \left[ \, {\pi^2 f_i^p \over (1-t)^i}\,+\, \int_1^\infty {F_i^p(z)\over (z-t)^i} \, dz \, \right]
\label{cf1pdr}
\ee


\section*{Appendix C: {\boldmath $\alpha_s$}-corrections to the moments}

Numerical values of the perturbative corrections to the moments ${{\bar M}^{(k),p}_n/{\bar M}^{(0),p}_n}$
and the $\alpha_s$-correction to the condensate contribution ${{\bar M}^{(O2)(1),p}_n/{\bar M}^{(O2),p}_n}$
are given in the Table 1 for $Q^2/(4{\bar m}^2)=0,1,2$ and $n=2-30$. The coefficient functions
 ${\bar M}^{p}_n$ are defined in $\ov{\rm MS}$ scheme according to (\ref{mpexb}, \ref{mommsb}).
 Remind, that the expansion (\ref{mpexb}) goes by $a({\bar m}^2)$. If one takes the QCD coupling
 at another scale, the function ${\bar M}^{(2), p}_n$ must be changed according to (\ref{m2shift}).

\begin{table}[p]
\caption{$\alpha_s$-corrections to PS-moments in $\ov{\rm MS}$ scheme}
$$
\ba{|c|ccc|ccc|ccc|}\hline
 Q^2 &\multicolumn{3}{|c|}{{\bar M}^{(1),p}_n/{\bar M}^{(0),p}_n} &
  \multicolumn{3}{|c|}{{\bar M}^{(2),p}_n/{\bar M}^{(0),p}_n} &
  \multicolumn{3}{|c|}{{\bar M}^{(O2)(1),p}_n/{\bar M}^{(O2),p}_n} \\ \cline{2-10}
 n & 0 & 4{\bar m}^2 & 8{\bar m}^2 & 0 & 4{\bar m}^2 & 8{\bar m}^2& 0 & 4{\bar m}^2 & 8{\bar m}^2 \\ \hline
2 &2.357 &0.825 &-0.176& -1.132 &-11.89 &-16.24 &1.489 &3.506 &3.617\\
3 &3.87 &2.976 &2.064& 12.85 &-0.466 &-8.633 &-1.58 &3.161 &3.731\\
4 &3.976 &3.894 &3.179& 19.28 &9.077 &-0.018 &\infty &2.486 &3.611\\
5 &3.508 &4.309 &3.831& 20.86 &15.63 &7.025 &3.91 &1.47 &3.32\\
6 &2.716 &4.442 &4.226& 19.8 &19.83 &12.51 &0.34 &-0.021 &2.88\\
7 &1.71 &4.391 &4.454& 17.56 &22.25 &16.68 &-1.955 &-2.424 &2.291\\
8 &0.55 &4.209 &4.562& 15.11 &23.33 &19.75 &-3.979 &-7.447 &1.536\\
9 &-0.727 &3.927 &4.581& 13.18 &23.42 &21.93 &-5.923 &-31.87 &0.567\\
10 &-2.098 &3.568 &4.529& 12.33 &22.76 &23.37 &-7.847 &31.4 &-0.713\\
11 &-3.545 &3.146 &4.42& 13. &21.58 &24.19 &-9.773 &11.7 &-2.508\\
12 &-5.057 &2.671 &4.264& 15.54 &20.04 &24.51 &-11.71 &6.907 &-5.312\\
13 &-6.623 &2.152 &4.068& 20.24 &18.29 &24.41 &-13.66 &4.376 &-10.66\\
14 &-8.237 &1.594 &3.837& 27.37 &16.44 &23.97 &-15.63 &2.6 &-26.79\\
15 &-9.892 &1.003 &3.577& 37.13 &14.6 &23.26 &-17.62 &1.16 &338.4\\
16 &-11.58 &0.383 &3.29& 49.72 &12.85 &22.33 &-19.62 &-0.105 &27.93\\
17 &-13.31 &-0.264 &2.98& 65.29 &11.27 &21.23 &-21.64 &-1.269 &15.2\\
18 &-15.07 &-0.934 &2.649& 84.01 &9.935 &20.01 &-23.67 &-2.373 &10.4\\
19 &-16.85 &-1.626 &2.3& 106. &8.893 &18.7 &-25.72 &-3.439 &7.707\\
20 &-18.66 &-2.336 &1.933& 131.4 &8.205 &17.34 &-27.78 &-4.479 &5.863\\
21 &-20.49 &-3.064 &1.551& 160.3 &7.917 &15.97 &-29.85 &-5.504 &4.447\\
22 &-22.34 &-3.809 &1.155& 192.8 &8.075 &14.6 &-31.93 &-6.519 &3.274\\
23 &-24.21 &-4.568 &0.745& 229. &8.718 &13.27 &-34.03 &-7.527 &2.252\\
24 &-26.1 &-5.341 &0.324& 268.9 &9.884 &12. &-36.14 &-8.531 &1.327\\
25 &-28.01 &-6.128 &-0.109& 312.8 &11.61 &10.81 &-38.26 &-9.533 &0.47\\
26 &-29.93 &-6.926 &-0.552& 360.5 &13.91 &9.728 &-40.38 &-10.53 &-0.339\\
27 &-31.87 &-7.735 &-1.005& 412.3 &16.84 &8.76 &-42.52 &-11.54 &-1.114\\
28 &-33.82 &-8.555 &-1.467& 468.1 &20.41 &7.929 &-44.66 &-12.54 &-1.864\\
29 &-35.78 &-9.385 &-1.938& 528.1 &24.65 &7.251 &-46.82 &-13.55 &-2.595\\
30 &-37.76 &-10.22 &-2.417& 592.2 &29.58 &6.74 &-48.98 &-14.55 &-3.311\\
\hline
\ea
$$
\end{table}

\end{document}